\documentstyle[11pt,aaspp4]{article}
\def\etal{et al.\ }
\def\deg{$^{\circ}$}
\begin{document}
\phantom{~}
\vspace{2cm}
\title{Bi-Directional Relativistic Jets of the Radio Galaxy 1946+708: \\
Constraints on the Hubble Constant}
\author{G. B. Taylor }
\affil{NRAO, Socorro, NM 87801; gtaylor@nrao.edu}

\and 
\medskip

\author{R. C. Vermeulen }
\affil{NFRA, Dwingeloo, The Netherlands; rcv@nfra.nl}
 
\vspace{1cm}
\hfil{~~~~~~~~~~Astrophysical Journal Letters, in press}\hfil
\vspace{1cm}

\begin{abstract}

We present measurements of bi-directional motions in the jets of
the radio galaxy 1946+708 at $z=0.101$. This is a Compact
Symmetric Object with striking S-symmetry. Sensitive 15 GHz
observations reveal a compact component at the center of symmetry with
a strongly inverted spectrum, that we identify as the core. From five
4.9 GHz observations spread over 4 years we have determined the
velocities of four compact jet components. If simple kinematic
models can be applied then the inclination of the source and the bulk
jet velocity can be directly determined for any assumed value of the
Hubble constant. Conversely, the measurements already place
constraints on the Hubble constant, and we show how further
observations of 1946+708 can yield an increasingly accurate
determination of $H_0$.

\end{abstract}


\section{Introduction}

A direct measure of the distance to an object can be obtained by
observing (angular) motion in it, if the intrinsic (linear) velocity
can be ascertained independently. Lynden-Bell (1977) first suggested
that Hubble's constant could be determined from observations of
superluminal extragalactic radio sources. While he assumed a
light-echo model,
the idea can be
generalized to the now commonly accepted relativistic jet model (e.g.\ 
Marscher \& Broderick 1982).
Here, we will make use of the additional constraints obtained by
observing relativistic motion of a pair of knots in anti-parallel
jets; we believe this is the first such detection in an AGN.

The radio source 1946+708 is identified with an $m_{\rm v}=18$ galaxy
at a redshift $z=0.101$ (Stickel \&\ K\"uhr 1993).
This
source is one of a family of Compact Symmetric Objects (CSOs)
comprising $\sim$5\% of sources in complete flux limited samples
selected at high frequencies (Readhead \etal 1996; Taylor \etal
1996a).  The CSOs are defined as sources less than 1 kpc in size
having radio emission on both sides of the central engine that is
thought to be relatively free of beaming effects (Wilkinson \etal
1994).  Nearly all CSOs have two steep-spectrum hot-spots and/or lobes
and most have an inverted or flat-spectrum core (Taylor \etal 1996a).
Preliminary measurements of the jet motions based on
the first two epochs at 5 GHz were discussed by Taylor, Vermeulen, \&
Pearson (1995).

\section{Observations and Data Reduction}

The first VLBI observation of 1946+708 was made on 1992 Sep 24 using
multiple snapshots with a global array of 15 antennas at 4.9 GHz
(Taylor \etal 1994).  The second epoch observation was made on 1994
Sep 15 in a similar fashion, although with only 12 antennas. The
telescopes used include those in the European VLBI Network, the Very
Long Baseline Array (VLBA) operated by NRAO\footnote{The National
Radio Astronomy Observatory is operated by Associated Universities,
Inc., under cooperative agreement with the National Science
Foundation}, the Very Large Array$^1$, the NRAO 140-foot$^1$ and the
Haystack Observatory.  Three further 4.9 GHz epochs were taken using the VLBA,
at epochs 1995 Mar 22, 1995 Sep 03, and 1996 Aug 18.
In addition
we also present 8.4 and 15 GHz
VLBA
observations made on
1996 July 7.
The calibration,
fringe-fitting, and mapping was performed following the procedures
described by Taylor \etal (1994).

In Fig.~1 we show nearly contemporaneous observations at 5, 8 and 15
GHz; the 8 GHz image has the greatest sensitivity.
%
Model-fitting of Gaussian components to the self-calibrated visibility
data was performed on each 5 GHz epoch using Difmap (Shepherd, Pearson
\& Taylor 1994, 1995).  The shapes of the components were fixed after
fitting to the first epoch, and in subsequent epochs each component
was allowed only to move, and to vary in flux density in order to fit
the independently self-calibrated visibility data.  The reduced
$\chi^2$ of the fit between the model and data is 1.06, 1.04, 0.94,
1.07, and 0.97 for epochs 1--5 respectively.  The errors in the
component positions were 
determined as
the shifts
that result
in a significant
(2\%) increase in the reduced $\chi^2$ of the fit after all other
components had been allowed to reconverge.

\section{Discussion}

\subsection{Location of the core and hot spots}

In Fig.~2 we plot the integrated spectrum of 1946+708, from
measurements using the VLA and the Owens-Valley millimeter array. Also
shown are the individual spectra of components C (inverted) and NHS
(steep), from our VLBA observations and a 1.3 GHz VLBA observation by
Conway \& Taylor (1997). The striking S-symmetry (Fig.~1) strongly
reinforces the idea that the centrally located compact inverted
spectrum component, C, is the center of activity.  We further believe
that the outer components NHS and SHS are genuine terminal hot spots;
a 21 cm VLA image shows 1946+708 to be unresolved in a 1.7\arcsec\ beam
with no extended component stronger than 0.11 mJy/beam (Taylor \etal
1996b).

We
measure a formally insignificant expansion rate between the two hot
spots of 0.22 $\pm$ 0.3 mas/yr.  We have also measured
the separation rate between the strong northern hot spot and the core
at 15 GHz to be 0.03 $\pm$ 0.03 mas in 1.29 years.  This latter
measurement gives a 3$\sigma$ upper limit on the advance speed of the
northern hot spot of $<0.4 h^{-1} c$, and implies an age for the
source of $>$ 200 years.

Unfortunately, the core component is too weak in the 5 GHz
images to use as a reference in aligning the epochs.  Therefore,
we have used the
strong northern hotspot as the reference.
Our conclusion, below, of the existence of bi-directional
motion in the jets is not critically dependent on this choice
and
can be avoided only by assuming that the core is
somewhere in the southern part of the source (for example, if S2 were
stationary), but this would require a significant velocity for
the northern hot spot (NHS) and a jet component (C) with bizarre
properties -- both of which seem
unlikely.  The positions of all components are plotted with respect to
the midpoint between the hot
spots.
This midpoint is $<0.2$ mas from the intersection point of the
line connecting N5 and S5 with the line connecting N2 and S2, and is
$<0.3$ mas from the core component visible in the 15 GHz image
(Fig.~1).

The trajectories of the two best defined pairs, N2/S2 and N5/S5, are
shown in Fig.~3.  To within the measurement errors these trajectories
can be fit with a straight line on the sky.  In Fig.~4 we show the
motion of each component projected along the fitted trajectories.  The
slope of this line corresponds to the velocity of the component.
Based on these observations no acceleration or deceleration of components is
required.

\subsection {Kinematics in 1946+708; Constraints on the Hubble Constant}

For simultaneously ejected components moving in opposite directions at
an angle $\theta$ to the line of sight at a velocity $\beta$, it
follows directly from the light travel time difference that the ratio
of apparent projected distances from the origin ($d_{\rm a}$ for the
approaching side, $d_{\rm r}$ for the receding side) as well as the
ratio of apparent motions (approaching: $\mu_{\rm a}$, receding:
$\mu_{\rm r}$) is given at any time by
\begin{equation}
{{\mu_{\rm a}}\over{\mu_{\rm r}}} = {{d_{\rm a}}\over{d_{\rm r}}} = \Biggr({{1+\beta\cos \theta}\over{1-\beta\cos \theta}}\Biggl)\,.
\end{equation}

There is a similar relationship for the ratio between the
flux density on the approaching side, $S_a$ to that on the receding
side, $S_r$, after including the effects of Doppler beaming: 
\begin{equation}
{{S_a}\over{S_r}} = \Biggl({{1+\beta\cos \theta}\over{1-\beta\cos \theta}}\Biggr)^{k-\alpha}\,,
\end{equation}
where $\alpha$ is the spectral index ($S\propto\nu^{\alpha}$), and
$k=2$ for a continuous jet or $k=3$ for discrete jet components (but
see e.g.\ Lind \&\ Blandford 1985).

We can apply the above equations to derive the product
$\beta\cos\theta$ in 1946+708,
assuming that C is the core
and that pairs of identical
components N5/S5 and N2/S2 were ejected simultaneously.
Components N5/S5 are still rather close to
the core, 
and there is evidence for
systematic
position
errors due to blending of the features,
so
we will not
yet
analyze N5/S5.
Using
the proper motion ratio $\mu_{\rm N2} / \mu_{\rm S2}=2.2\pm0.9$ yields
$\beta\cos\theta = 0.38\pm0.13$. However, this derivation is subject
to substantial additional systematic errors due to the uncertainty in
pinpointing the stationary reference point. The flux density ratio
$S_{\rm N2} / S_{\rm S2}=1.86 \pm 0.11 $ indicates $\beta\cos\theta =
0.09 \pm 0.03$, for $k=3$, and $\beta\cos\theta = 0.12\pm0.03$, for
$k=2$, with a spectral index of $\alpha=-0.6$, estimated from the
multi-frequency images.  This is subject of course to the further
assumption that the emitted fluxes are still identical, and it is
somewhat remarkable that the result is close to what we believe to be
the most accurate estimate, $\beta\cos\theta = 0.16 \pm 0.01$, which
follows from $d_{\rm N2}/d_{\rm S2}=1.38\pm0.03$.
We plot this constraint
on Figure 5.

Another constraint on the two parameters $\beta$ and $\theta$ can be
obtained from the separation rate $\mu_{\rm sep} = |\mu_{\rm a}| +
|\mu_{\rm r}|$, which, unlike $\mu_{\rm a} / \mu_{\rm r}$, is not subject to
the uncertainty in the reference point. From geometry and the
conversion of angular to linear velocity we have:
\begin{equation}
v_{sep} = {\mu_{sep}\,D_a\,(1+z)} = {{2 \beta c \sin \theta}\over{(1 - \beta^2\cos^2 \theta)}}\,,
\end{equation}
where $v_{sep}$ is the projected separation velocity, $D_a$ is the angular size distance to the source, and $z$ is the
redshift. We will take $q_0 = 0.5$ in Friedmann cosmology; this choice
is unimportant given the low redshift $z=0.101$ of 1946+708. We
measure $\mu_{\rm sep N2-S2} = 0.17 \pm 0.028$ mas/yr which gives $v_{\rm sep N2-S2} = (0.74 \pm 0.12)\, h^{-1} c$, with $H_0=100h$
km s$^{-1}$ Mpc$^{-1}$. The resultant locus of $\beta$ and $\theta$ is
illustrated in Figure 5 for two choices: $h=1$ and $h=0.37$.

For $h=1$, the intersection with $\beta\cos\theta$ is already at a
substantial angle to the line-of-sight ($\theta \sim 65$\deg) and a
moderate value of $\beta \sim 0.4$. Smaller angles and smaller jet
velocities (limit: $\beta\ge0.15$ because $\theta\ge0$\deg) would need
an implausibly high Hubble constant. On the other hand, the fact that
$\beta < 1$ not only gives the weak limit $\theta< 81$\deg, but also
implies that $h>0.37$ ! The $\beta\cos\theta$ area in Fig 5 is nearly
vertical for all plausible values of $H_0$, meaning that $\theta$ is
constrained to the narrow range 65--80\deg, while the allowed values
of $\beta$ and $h$ are roughly inversely proportional. Thus, for
relativistic jets with $\beta\ge0.9$, as are frequently assumed to
exist in high luminosity AGN, values of $H_0$ towards the lower end of
the commonly discussed range are much more plausible than high values.

There are several additional methods which can in principle be used to
further constrain the jet Doppler factor or velocity, and hence improve our
determination of $H_0$.  These are: (1) flux density variability studies (Rees
1967); (2) comparison of predicted and observed inverse Compton X-ray
emission (Marscher \& Broderick 1982); (3) measurement of 
relativistic aberration (Unwin \& Wehrle 1992); and
(4) measurements of equipartition Doppler factors
(Readhead 1994).  The various parameters involved scale with
$H_0$ to different powers, effectively leading to different curves in a diagram
like Figure 5.  Thus, a combination of more methods can improve the 
determination of $H_0$, and we plan to pursue all of them.  

\subsection {Evolution of the Jet Components}


Assuming that the components N2 and S2 were ejected at the same time
and have kept a constant velocity, their age at the time of the first
epoch observations was 93 years, implying ejection in 1899. The age of
N5 and S5 would be 8 years, with an ejection date of 1984.
At $\theta\ge65$\deg,
the observed 
flux densities of the jet
components in 1946+708 are only mildly Doppler boosted.  Therefore
they have a greater intrinsic surface brightness than the jets found
in larger radio galaxies or even in typical parsec-scale core-jet
sources (Taylor \etal 1994).



The source shows significant curvature, and at
$\theta\ge65$\deg\ this must be largely intrinsic. If
the
components 
are indeed
moving on curved tracks
we
expect a discrepancy between the geometry derived from the distance
ratio $d_{\rm a}/d_{\rm r}$, which depends on the time-integral angle
since ejection, and the motion ratio $\mu_{\rm a}/\mu_{\rm r}$, which
reflects the angle only during the monitoring interval. There is some
suggestion from our analysis above that knots N2 and S2 might indeed
be in the process of curving into and away from the line of sight,
respectively. On the other hand, the current monitoring series does
not rule out that the knots might be moving ballistically. In that
case, one might attribute the overall curvature to precession
of the central engine.
The
current data do not warrant fitting of a
model,
but the precession period
would be quite
short ($<$200 years), much less than expected for a stable binary
black hole in the model proposed by Begelman, Blandford \& Rees
(1980). Prolonged astrometric monitoring will surely be very
illuminating.

\section {Conclusions}

Components in the parsec-scale jet and counterjet in 1946+708 are
observed to move away from the center of activity.  Pairing components
up under the assumption of simultaneous ejections
gives reasonable agreement between
arm length,
flux
density,
and velocity ratios.  These relations also allow us
to constrain Hubble's constant to $H_{0} \ge 37$.  Future
measurements, especially those carried out at higher frequencies, will
further elucidate the source geometry and improve in accuracy the
constraints on the motions and thus on the Hubble constant linearly
with time.  We are also in the process of examining other CSOs
for bi-directional motions.  If enough of these twin-jet
systems can be found over a range in redshift then they might
eventually provide a direct determination of $q_0$ as well.

\acknowledgments
We thank Tim Pearson for his encouragement and involvement 
in the early stages of this work.  We are grateful to an anonymous referee for
helpful comments.
We thank the staffs at the observatories and the staff of the
JPL/Caltech Block II Correlator and the VLBA Correlator for their
assistance.  This work was supported in part by the NSF under grants
AST-9117100 and AST-9420018.

\clearpage

\begin{figure}
\vspace{15cm}
\includegraphics{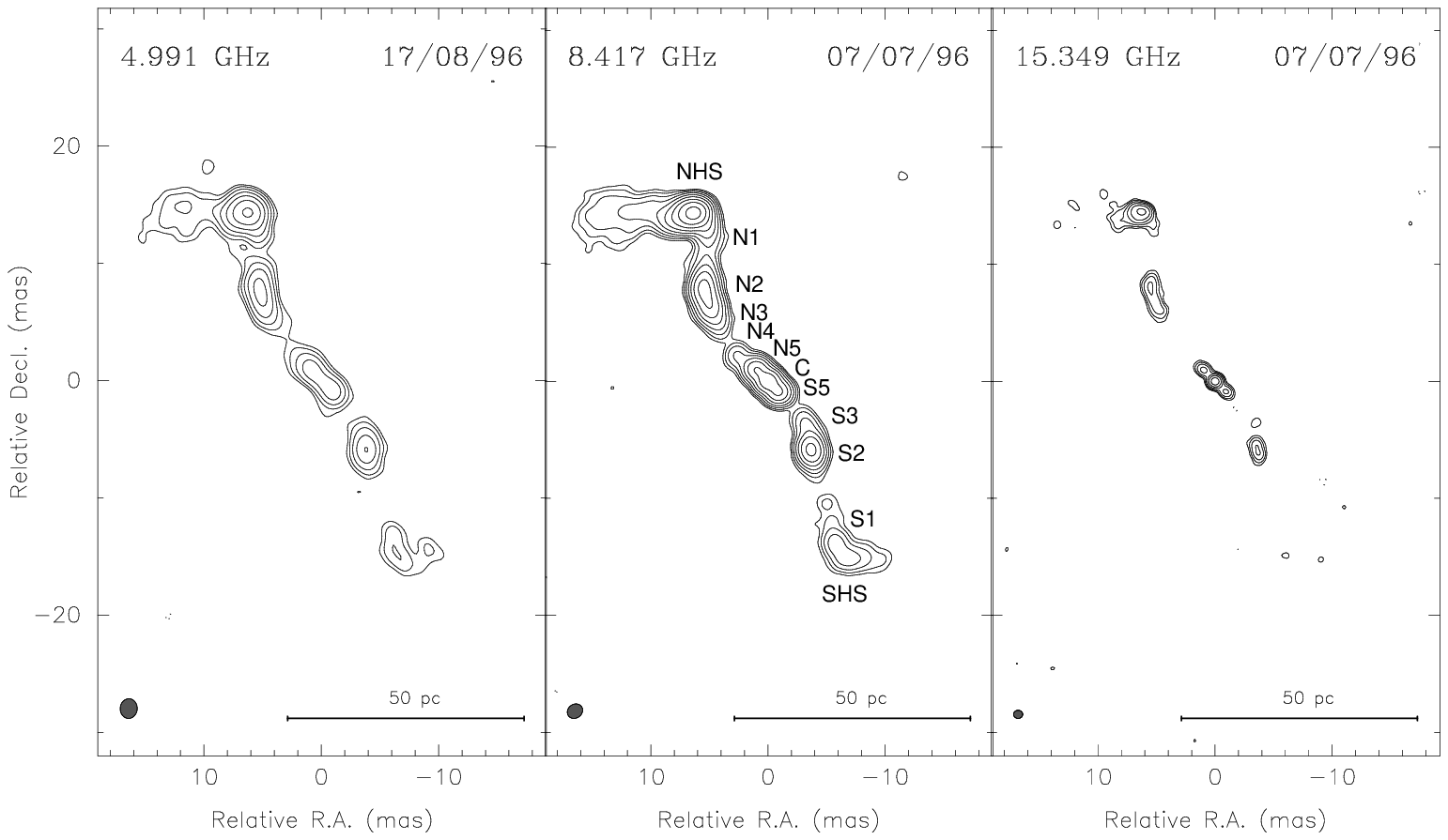}
\figcaption{Nearly contemporaneous VLBA observations of 1946+708 at 5, 
8 and 15 GHz.
The synthesized beam is drawn in the lower left-hand corner of
each plot and has 
dimensions: 1.68 $\times$ 1.45 mas in position angle $-$3.4\deg\ at 5
GHz; 1.37 $\times$ 1.18 mas in position angle $-$55\deg\ at 8
GHz; and 0.78 $\times$ 0.7 mas in position angle 90\deg\ at 15
GHz.
Contours are drawn logarithmically at factor 2 intervals with 
the first contour at 2, 0.25, and 0.75 mJy/beam at 5, 8 and 15 GHz
respectively.  The components labeled NHS and SHS in the 8 GHz image
are the northern
and southern hot spots respectively.  The component labeled ``C''
we identify as the core.  }
\end{figure}
\clearpage

\begin{figure}
\vspace{15cm}
\includegraphics{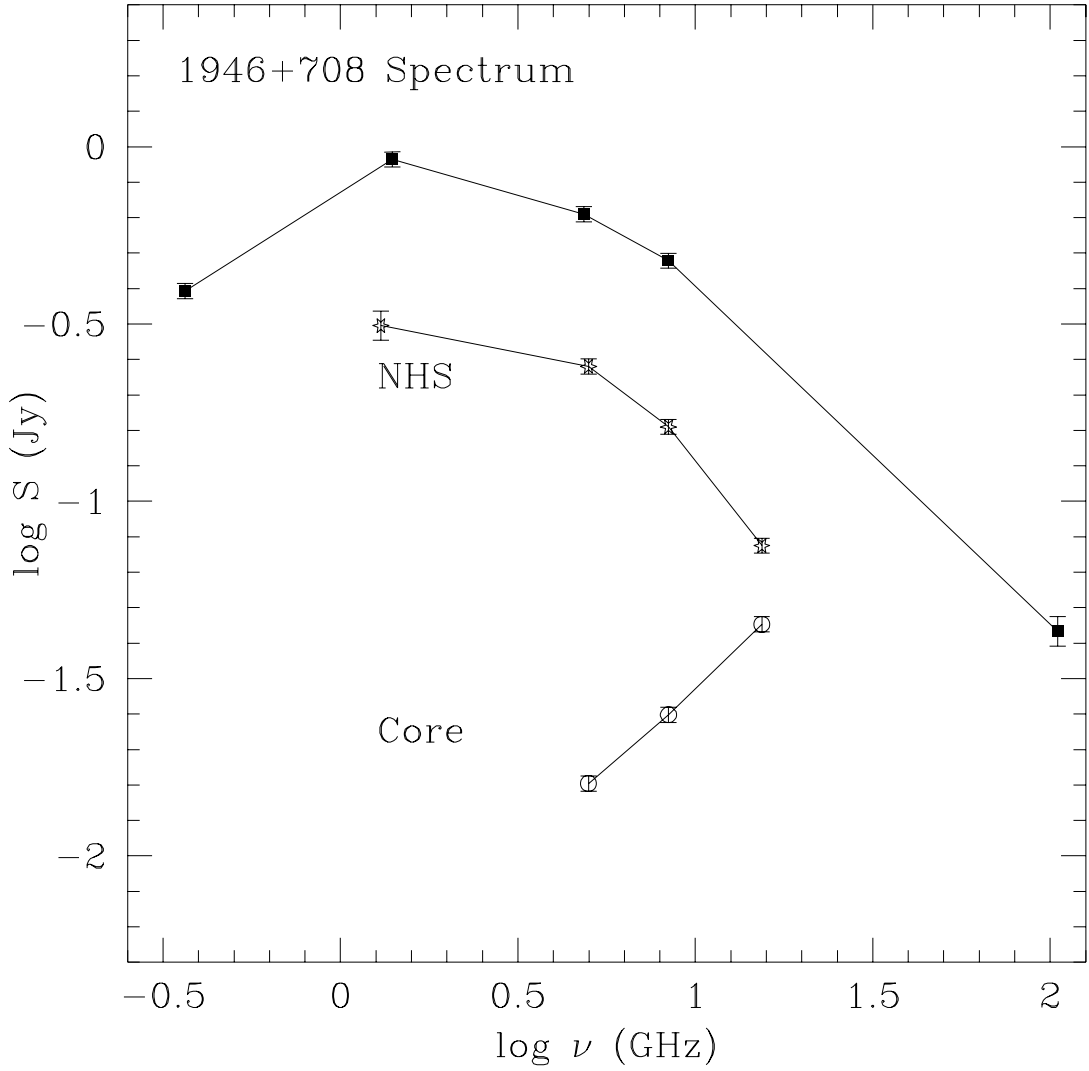}
\figcaption{The spectrum of 1946+708.  Total flux densities (at 
various epochs, from Taylor \etal 1996b) are shown plotted as filled
squares.  Flux densities for the northern hot spot (stars) and core
component (circles) have been derived from modelfits to the VLBA 1996
epoch data except for the 1.3 GHz measurement which comes from VLBA data at
epoch 1995 Mar 22 as described in Conway \& Taylor (1997).}
\end{figure}
\clearpage

\begin{figure}
\vspace{15cm}
\includegraphics{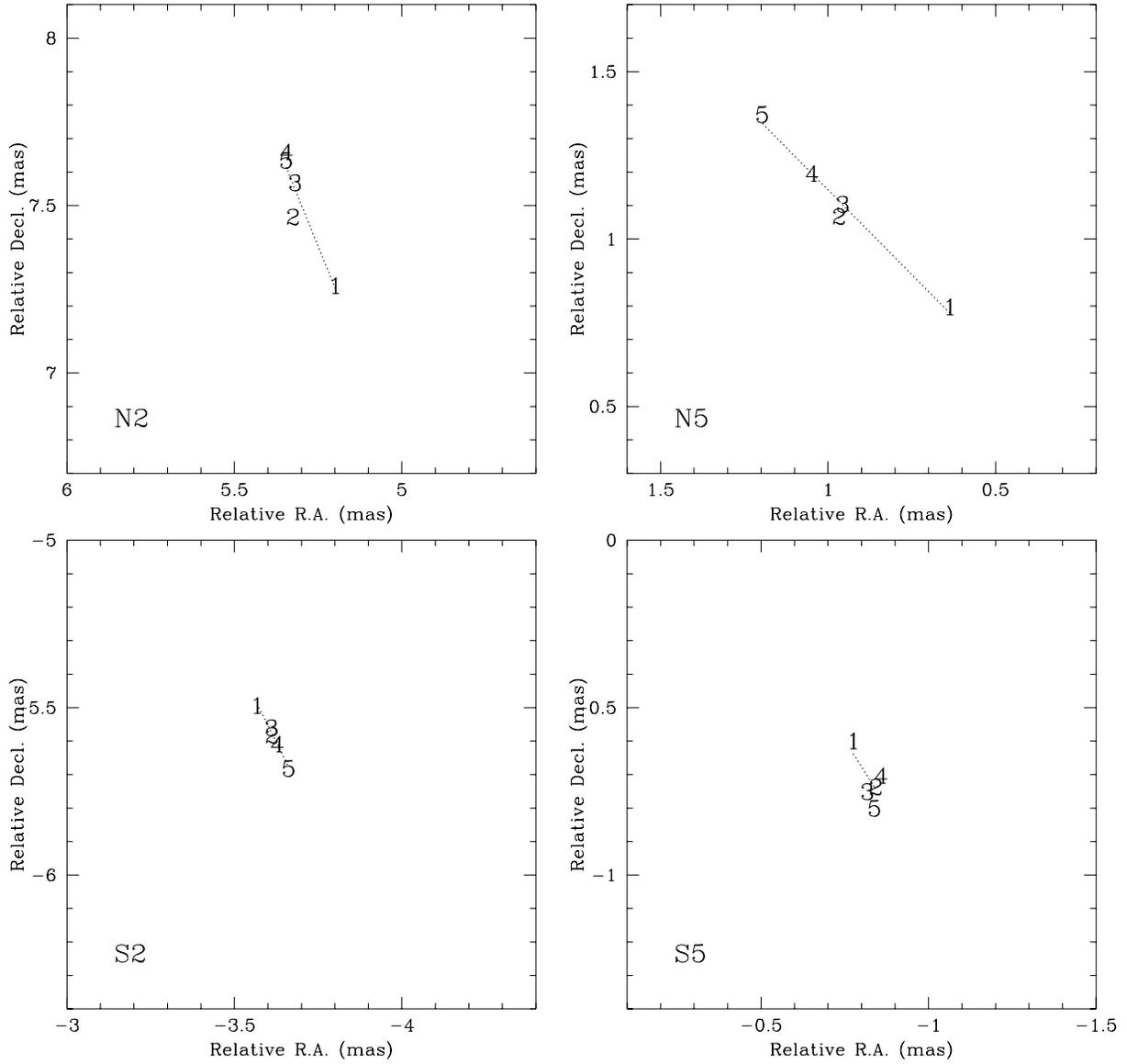}
\figcaption{Positions of the components N2, S2, N5 and S5 relative to the
source center.  The best-fit straight-line motions are shown as a
dotted line.  The
numbers correspond to the position of the component at epochs 1 = 1992
Sep 24, 2 = 1994 Sep 15, 3 = 1995 Mar 22, 4 = 1995 Sep 3, and 5 = 1996
Aug 18.}
\end{figure}
\clearpage

\begin{figure}
\vspace{15cm}
\includegraphics{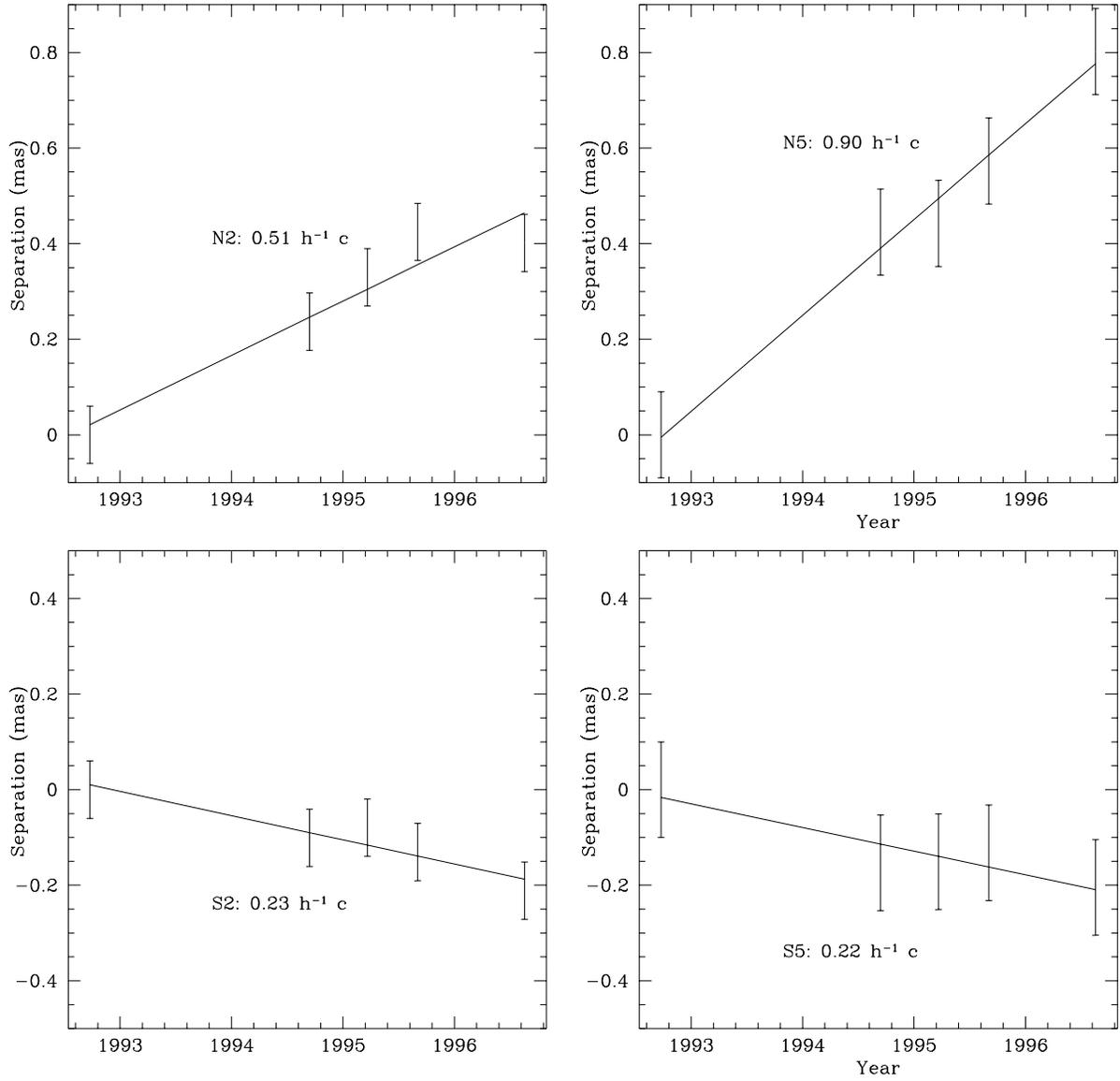}
\figcaption{Velocities of the components N2, S2, N5 and S5 fitted to the  
projected positions along the straight line which best fits the 
measured positions (see Fig.~3).
The zero point is taken to be the position of the component at
the first epoch.  The southern components are plotted with negative
separations to reflect their near 180\deg\ difference in direction.}
\end{figure}
\clearpage

\clearpage
\begin{figure}
\vspace{15cm}
\includegraphics{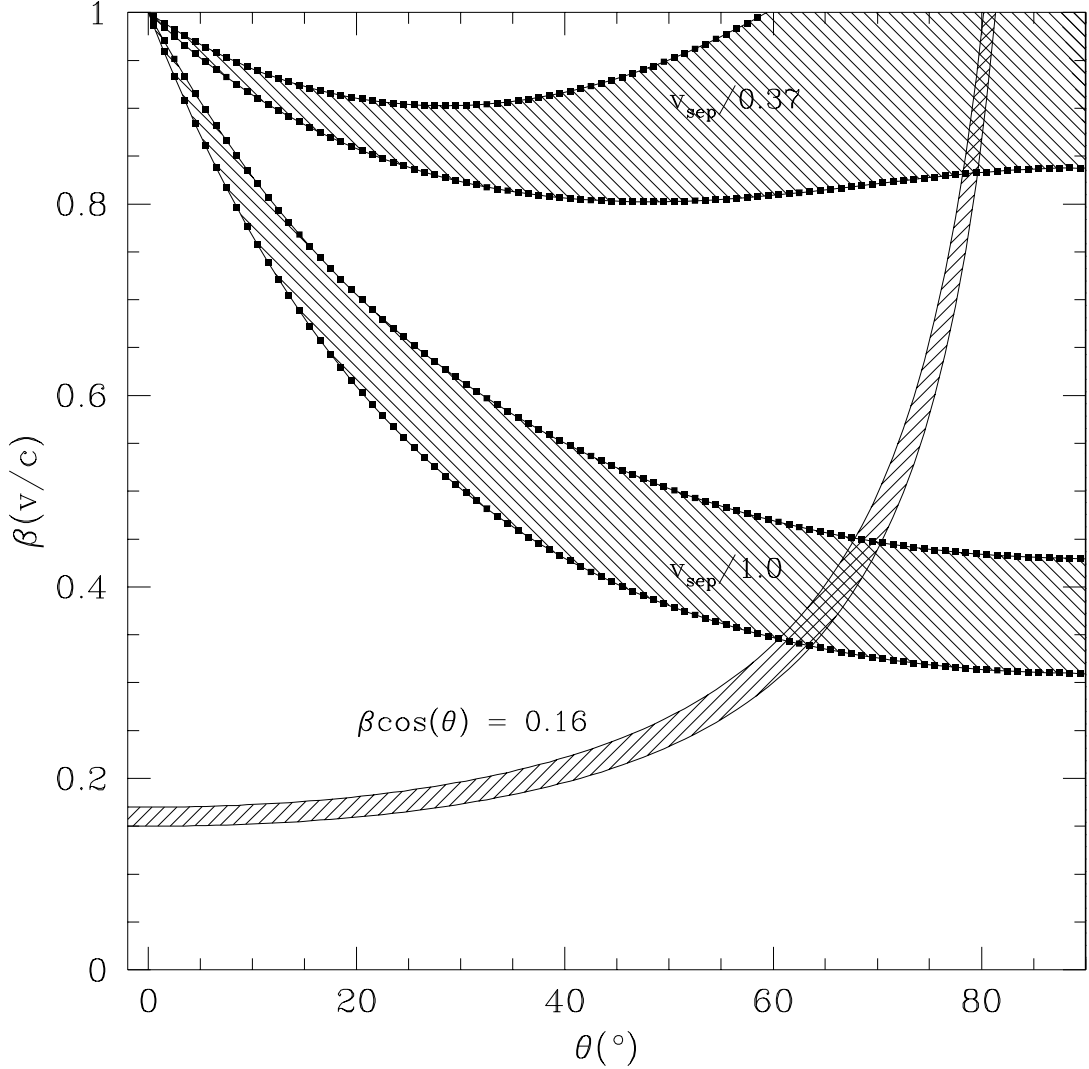}
\figcaption{The jet velocity ($\beta$) plotted against the inclination
of the source ($\theta$) measured from the line-of-sight to the jet
axis.  The solid lines represent the constraint $\beta \cos\theta =
0.16 \pm 0.01$ from the arm-length ratio of components N2 and S2
(Eq.~1).  The heavy dashed lines show the constraint from the observed
separation velocity, $h^{-1} v_{sep}$, for N2 and S2 with $h$ =
0.37 $\pm$ 0.06, and 1.0 $\pm$ 0.16 where $h = H_0/100$ km s$^{-1}$ Mpc$^{-1}$
(Eq.~3).  If $\beta = 1$ then $h = 0.37 \pm 0.06$ where the
uncertainty in $h$ stems from uncertainty in the
measurement of $v_{sep}$.
}
\end{figure}
\clearpage

\end{document}